\documentclass[final,authoryear,5p,times,twocolumn]{elsarticle}
\pdfoutput=1

\usepackage{graphicx}

\usepackage{amssymb}

\usepackage[pdftex,pdfpagemode={UseOutlines},bookmarks,bookmarksopen,colorlinks,linkcolor={blue},citecolor={green},urlcolor={red}]{hyperref}
\usepackage{hypernat}

\journal{Astronomy \& Computing}

\usepackage{upquote}

\usepackage{upgreek}

\newcommand*\secref[1]{Sect.~\ref{#1}}
\newcommand*\figref[1]{Fig.~\ref{#1}}
\newcommand*\tabref[1]{Table~\ref{#1}}

\begin{document}

\begin{frontmatter}



\title{Learning from 25
  years of the extensible \emph{N}-Dimensional Data Format}


\author[cornell]{Tim Jenness\corref{cor1}}
\ead{tjenness@cornell.edu}
\author[jac]{David S. Berry}
\author[jac]{Malcolm J.\ Currie}
\author[durham]{Peter W.\ Draper}
\author[noao]{Frossie Economou}
\author[glasgow]{Norman Gray}
\author[ral]{Brian McIlwrath}
\author[aao]{Keith Shortridge}
\author[bristol]{Mark B.\ Taylor}
\author[ral]{Patrick T.\ Wallace}
\author[ral]{Rodney F.\ Warren-Smith}

\cortext[cor1]{Corresponding author}

\address[cornell]{Department of Astronomy, Cornell University, Ithaca,
  NY 14853, USA}
\address[jac]{Joint Astronomy Centre, 660 N.\ A`oh\=ok\=u Place, Hilo, HI
  96720, USA}
\address[durham]{Department of Physics, Institute for Computational Cosmology, University of Durham, South Road, Durham DH1 3LE, UK}
\address[noao]{LSST Project Office, 933 N.\ Cherry Ave, Tucson, AZ 85721, USA}
\address[glasgow]{SUPA School of Physics \& Astronomy, University of Glasgow, Glasgow G12 8QQ, UK}
\address[ral]{RAL Space, STFC Rutherford Appleton Laboratory, Harwell Oxford, Didcot, Oxfordshire OX11 0QX, UK}
\address[aao]{Australian Astronomical Observatory, 105 Delhi Rd, North
Ryde, NSW 2113, Australia}
\address[bristol]{H.\ H.\ Wills Physics Laboratory, Bristol University, Tyndall Avenue, Bristol, UK}

\begin{abstract}

The extensible \emph{N}-Dimensional Data Format (NDF) was designed and
developed in the late 1980s to provide a data model suitable for use
in a variety of astronomy data processing applications supported by
the UK Starlink Project. Starlink applications were used extensively, primarily in
the UK astronomical community, and form the basis of a number of
advanced data reduction pipelines today. This paper provides an
overview of the historical drivers for the development of NDF and the
lessons learned from using a defined hierarchical data model for many
years in data reduction software, data pipelines and in data
acquisition systems.

\end{abstract}


\end{frontmatter}


\newcommand{\mnras}{MNRAS}
\newcommand{\aap}{A\&A}
\newcommand{\aaps}{A\&AS}
\newcommand{\pasp}{PASP}
\newcommand{\apj}{ApJ}
\newcommand{\apjs}{ApJS}
\newcommand{\qjras}{QJRAS}
\newcommand{\an}{Astron.\ Nach.}
\newcommand{\ijimw}{Int.\ J.\ Infrared \& Millimeter Waves}
\newcommand{\procspie}{Proc.\ SPIE}
\newcommand{\aspconf}{ASP Conf. Ser.}


\newcommand{\KAPPA}{\textsc{kappa}}
\newcommand{\gaia}{\textsc{gaia}}
\newcommand{\figaro}{\textsc{figaro}}
\newcommand{\ccdpack}{\textsc{ccdpack}}
\newcommand{\smurf}{\textsc{smurf}}
\newcommand{\surf}{\textsc{surf}}
\newcommand{\asterix}{\textsc{asterix}}
\newcommand{\specdre}{\textsc{specdre}}
\newcommand{\iras}{\textsc{iras90}}
\newcommand{\treeview}{\textsc{treeview}}
\newcommand{\splat}{\textsc{splat}}
\newcommand{\catpac}{\textsc{catpac}}
\newcommand{\CFITSIO}{\textsc{cfitsio}}
\newcommand{\fitstondf}{\textsc{fits}{\footnotesize{2}}\textsc{ndf}}

\newcommand{\ascl}[1]{\href{http://www.ascl.net/#1}{ascl:#1}}

\section{Introduction}
\label{sec:intro}

There is a renewed interest in file-format choices for astronomy
with discussions on the future of FITS
\citep{P90_adassxxiii,2014Thomas}, projects adopting or considering
HDF5 \citep{2012ASPC..461..283A,jenness_spie2014} and investigations into
general purpose image formats such as JPEG2000 \citep{2014Kitaeff}. These
discussions have provided an opportunity to consider existing
astronomy file formats that are not as widely known in the community
as FITS. Here we discuss the extensible \emph{N}-Dimensional
Data Format (NDF) developed by the Starlink Project
\citep{2000ASSL..250...93W} in the late 1980s
\citep{1988STARB...2...11C,SGP38} to bring order to the proliferation
of data models that were being adopted by applications using the Starlink
hierarchical file format.

In \secref{sec:hds} we discuss the genesis and features of the
Starlink file format, with the NDF data model itself being discussed
in \secref{sec:ndf}.  In \secref{sec:lessons} we discuss the positive
lessons learned from developing NDF (expanding on some earlier work
by \citet{P91_adassxxiii}) and we follow that in \secref{sec:improve},
by discussing the areas where NDF could be improved.
Following this, in \secref{sec:social} we address the social,
political and economic considerations of data formats and models. In
\ref{app:chaos} we provide examples of data models devised in the
mid-1980s, the development of which motivated the creation of NDF
and also directly influenced the NDF design. A timeline describing the
key developments relating to NDF is shown in \tabref{tab:timeline}.

In this paper the term ``data model'' refers to the organization,
naming and semantics of components in a hierarchy. The term
``file format'' means how the bytes are arranged on disk and
in this context refers to the use of the Hierarchical Data System
(HDS; see the next section). Historically, NDF, as implied by the use
of ``Data Format'' in the name itself, uses the term ``format'' to refer to
the data model in the sense of how the hierarchical items are arranged
or ``formatted'' (cf. text formatting), and not specifically referring
to the underlying file format. In this paper we use try to use
consistent modern terminology although in some cases there can be ambiguity,
and in \secref{sec:social} we use ``format'' to mean the all-encompassing
concept of NDF as a whole.

\section{Hierarchical Data System}
\label{sec:hds}

The Starlink Project, created in 1980
\citep{1980IrAJ...14..197E,1982MmSAI..53...55T,1982QJRAS..23..485D}, was set up
primarily to provide data reduction software and facilities to United
Kingdom astronomers. FITS
\citep{1979ipia.coll..445W,1981A&AS...44..363W} had recently been
developed and was adopted as a tape interchange format but there was a
requirement for a file format optimized for data reduction that all
Starlink applications could understand. Files
needed to be written to and read from efficiently with an emphasis on the
ability to group and modify related information
\citep{1981STARENT4}. This was many years before the NCSA developed
the Hierarchical Data Format \citep{HDF1,Folk2010} and it was decided
to develop a new file format. The resulting Starlink Data
System\footnote{from which the file extension of \texttt{.sdf}, for
  Starlink Data File, was chosen.} was first proposed in 1981 with the
first version being released in 1982 \citep[see
e.g.][]{1982QJRAS..23..485D,1991STARB...8....2L}. In 1983 the name was
changed to the Hierarchical Data System (HDS) to make the file format
benefits more explicit. HDS itself was in common usage within Starlink by 1986
\citep{1986BICDS..30...13L}. It was originally written in the BLISS
programming language on a VAX/VMS system and later rewritten in C and
ported to Unix.
It was, however, designed to be only callable from FORTRAN at this point.

Some key features of the HDS design are as follows.
\begin{itemize}
\item It provides a hierarchical organization of arbitrary structures, including the
  ability to store arrays of structures.
\item The hierarchy is self-describing and can be queried.
\item It gives the data author the ability to associate structures with an arbitrary data type.
\item Users can delete, copy or rename structures within a file.
\item It supports automatic byte swapping whilst using the native machine byte order
  for newly created output files.
\item VAX and IEEE floating-point formats are supported.
\item Automatic type conversion allows a programmer to request that,
  say, a data array of 32-bit integers is accessed as 64-bit floating
  point numbers.
\end{itemize}

\begin{table}
\caption{Time line for developments relating to NDF. Other
  developments in file format development are interspersed
  for reference.}
\label{tab:timeline}
\begin{tabular}{|l|l|}
\hline
1979 & \emph{Flexible Image Transport System announced.}\\
1981 & HDS proposed. \\
1982 & HDS Version 1 ready for testing. \\
1983 & \emph{IMAGE} model proposed. \\
1988 & \emph{NCSA release the Hierarchical Data Format.}\\
1988 & \emph{FITS tables standard.}\\
1988 & NDF standard data structures released.\\
1991 & HDS ported to Sun OS (HDS Version 3).\\
1998 & World Coordinate Objects added to NDF.\\
2001 & \emph{Definition of FITS standard.}\\
2002 & \emph{HDFv5 is released.}\\
2005 & Support files larger than 2GB (HDS Version 4).\\
2005 & Add official C interface to HDS.\\
2007 & Provenance added to NDF as extension.\\
2010 & \emph{Version 3 of the FITS standard.}\\
2013 & 64-bit integer type added to HDS.\\
\hline
\end{tabular}
\begin{center}
\end{center}
\end{table}

More recently HDS has been extended to support 64-bit file
offsets so that modern large datasets can be processed\footnote{Support for
  individual data  arrays containing more elements than can be counted
  in a 32-bit integer is not yet possible.}, and also the addition of
a native C-language interface and a 64-bit integer data type
\citep{P82_adassxxiii}. The primitive data types supported by HDS are
listed in \tabref{tab:hdstypes}.

\begin{table}
\caption{HDS basic data types. The unsigned types did not correspond
  to standard Fortran~77 data types and were included for
  compatibility with astronomy instrumentation. HDS supports both VAX
  and IEEE floating-point formats. The API code indicates the letter appended
  to function names to indicate the type they support. This convention is
  used for the \textsc{generic} templating system \citep{SUN7}.}
\label{tab:hdstypes}
\begin{center}
\begin{tabular}{lll}
\hline
Name of type & API Code & Data type \\ \hline
\texttt{\_BYTE} & b & Signed 8-bit integer \\
\texttt{\_UBYTE} & ub & Unsigned 8-bit integer \\
\texttt{\_WORD} & w & Signed 16-bit integer \\
\texttt{\_UWORD} & uw & Unsigned 16-bit integer \\
\texttt{\_INTEGER} & i & Signed 32-bit integer \\
\texttt{\_INT64} & k &Signed 64-bit integer \\
\texttt{\_LOGICAL} & l & Boolean \\
\texttt{\_REAL} & r & 32-bit float \\
\texttt{\_DOUBLE} & d & 64-bit float \\
\texttt{\_CHAR[$*$n]} & c & String of 8-bit characters \\
\hline
\end{tabular}
\end{center}
\end{table}

The advantage of HDS over flat file formats is that it allows many different kinds of data to
be stored in a consistent and logical fashion. It is also very
flexible, in that objects can be added or deleted whilst retaining the
logical structure.  HDS also provides portability of data, so that the
same data objects may be accessed from different types of computer
despite platform-specific details of byte-order and floating-point format.

\section{The \emph{N}-Dimensional Data Format}
\label{sec:ndf}

\begin{figure}[t]
\includegraphics[width=\columnwidth]{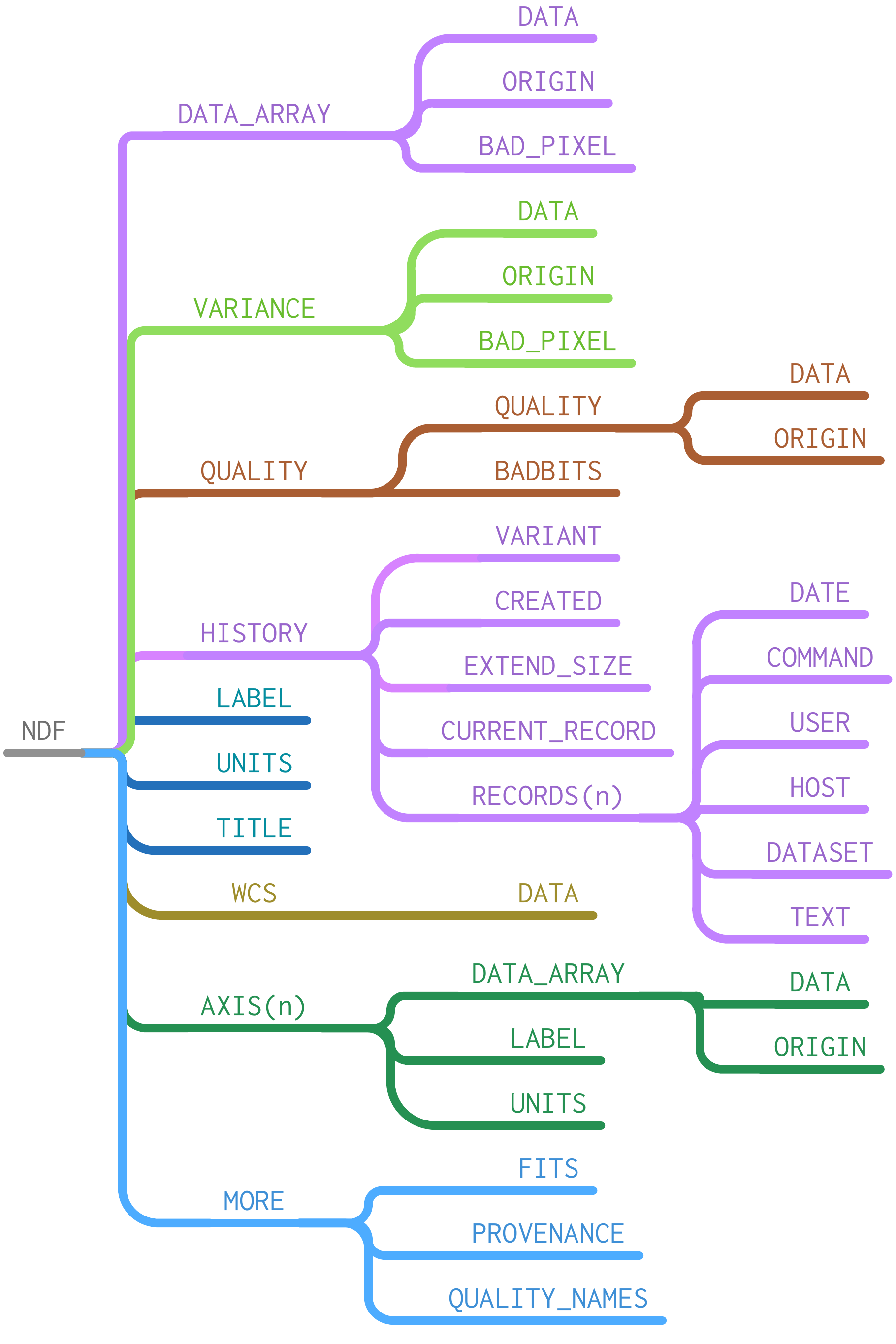}
\caption{Schematic of the NDF hierarchy. All components are optional
  except for \texttt{DATA\_ARRAY}. For an alternative visualisation
  see \citet{2002ASPC..281...20G}.}
\label{fig:ndf-structure}
\end{figure}

HDS files allowed people to arrange their data in the files however
they pleased and placed no constraints on the organization of the
structures or the semantics of the content. This resulted in
serious interoperability issues when moving files between applications
that nominally could read HDS files. Within the Starlink ecosystem
there were at least three prominent attempts at providing data
models, and these are discussed in detail in \ref{app:chaos}. The
models were: Wright-Giddings \emph{IMAGE} \citep{WrightGiddings1983}, \figaro\
\citep[][\ascl{1203.013}]{1993ASPC...52..219S} DST
and \asterix\
\citep[][\ascl{1403.023}]{1987JBIS...40..185P}.
The result was chaos.

Given the situation with competing schemes for using the hierarchy, it
became clear that a unified data model was required. A working group
was formed to analyze the competing models and come up with a
standard;\footnote{Throughout this document \emph{NDF standard} refers
to the written specification \citep{SGP38} developed by the working group
in collaboration with Starlink, and also the NDF library
that enforces the specification. Currently there is no standards body
working on the NDF specification and enhancements are added as
required through discussion on the developer mailing list and
 the existence of a single reference NDF library.} this work
was completed in the late 1980s \citep{1988STARB...2...11C,SGP38}.
After much debate, it was decided to develop a data model that included
the minimum structures to be useful for the general astronomer without
attempting to be everything to everybody, but designing in an extension
facility from the beginning. The resulting model was named the
extensible \emph{N}-Dimensional Data Format (NDF).  NDF combined some
features of the \emph{IMAGE} scheme, such as the use of
\texttt{DATA\_ARRAY}, and some features adopted from \asterix, from
which the \texttt{HISTORY} structure was adopted complete.  The
decision to recognize NDF structures  based solely on the presence of a
\texttt{DATA\_ARRAY} array was an important compromise as it allowed
data using the \emph{IMAGE} data model, which indirectly included data in the
Bulk Data Frame \citep[BDF;][]{1980SPIE..264...70P,SUN4}
format that had been converted previously, to be used immediately
in applications that had been ported to use the NDF library.

In the following sections we describe the core components of the NDF
data model to provide an overview of the NDF approach concerning what
is covered by the model and what is deliberately left out of the
model. An overview of the components of an NDF and how they relate to
each other is shown in \figref{fig:ndf-structure}. For more details
on NDF please see the detailed NDF design document
\citep[SGP/38;][]{SGP38} and library documentation
\citep[SUN/33;][]{SUN33}.

\subsection{Data arrays}

NDF supports the concept of a primary data array and an associated
variance array and quality mask. All the HDS numerical data types are available
but there is also support for complex numbers for the data and
variance components. Variance was selected as the error
component to reduce the computational overhead when propagating errors
during processing. All three components share the same
basic \texttt{ARRAY} structure which defines the dimensionality of the
array and allows for the concept of a pixel origin. The pixel origin
is used to specify where the NDF sits relative to a larger pixel
coordinate system by specifying the coordinates of the bottom-left
pixel. For example, when extracting a subset of data from
a bigger image, the pixel origin will record where the subset came
from. Also, when images are registered for mosaicking they are
resampled such that their origins share a common coordinate frame
resulting in the final mosaic being a simple pixel-by-pixel
combination. See \secref{sec:wcs} for a discussion on how the pixel
origin relates to other coordinate systems. The
\texttt{BAD\_PIXEL} flag is intended as a hint to application
software to allow simpler and faster algorithms to be used if it is known
that no bad values are present in the data array. Additionally, the flag
can be used to indicate that all values are to be used, thereby
disabling bad value handling and allowing the full range of a data
type. This was felt to be particularly important for the shorter
integer data types.

The quality mask uses the \texttt{ARRAY} type but includes an extra
level in the structure to support a bit mask. The \texttt{BADBITS}
mask can be used to enable or disable planes in the \texttt{QUALITY} array.

For applications that do not wish to support explicit quality
tracking, the NDF library supports automatic masking of the data and
variance arrays; this uses the quality mask in the input NDF, and sets
the corresponding value in the variance array to a magic value when
the data are mapped into memory. Unlike the \emph{IMAGE} scheme
or FITS, which allow the magic (or blank) value to be specified per
data array, NDF specifies the magic value to be used for each data type
covering both floating-point and integer representations, inheriting
the definition from the underlying HDS definition. Indeed,
\texttt{NaN} (defined by the IEEE floating-point model \citep{1985-754IEEE}) is not explicitly part of the standard and it is usually
best to convert \texttt{NaN} elements to the corresponding floating-point magic
value before further processing is applied. A single definition of the
magic value for each data type simplifies applications programming and
removes the need for the additional overhead of providing a value for
every primitive data array.

\texttt{NaN} was excluded from the standard as it was not
supported in VAX floating point \citep[see e.g.][]{660194} and the Starlink
software was not ported to machines supporting IEEE floating point
until the 1990s \citep[e.g.,][]{1991STARB...8...11C}. Unlike FITS, which
did not officially support floating point until 1990
\citep{1989FPFITS,1991BAAS...23..993S} when they were able to adopt
\texttt{NaN} as part of the standard, much software pre-existed
in the Starlink environment at this time and embodied
direct tests for magic values in data. Given the different semantics
when comparing data values for equality with \texttt{NaN}, it was decided to
continue with the magic value concept rather than try to support two
rather different representations of missing data simultaneously.

\subsubsection{Data compression}

The original NDF standard included the \emph{SCALED} data compression
variant, which is commonly required when representing floating-point numbers in
integer form; this is equivalent to \texttt{BSCALE} and \texttt{BZERO} in
FITS, although this variant was not implemented in the NDF library
until 2006 \citep{2008ASPC..394..650C}. In 2010 a lossless delta
compression system for integers was added to support raw SCUBA-2 data
\citep{2013MNRAS.430.2513H}. This was a new implementation of the
\texttt{slim} data compression
algorithm\footnote{\url{http://sourceforge.net/projects/slimdata/}}
developed for the Atacama Cosmology Telescope
\citep[][\ascl{1409.010}]{2004SPIE.5498....1F}.

\subsection{Character attributes}

Each NDF has three character attributes that can be specified: a
title, a data label and a data unit. These values can be accessed
through the library API without resorting to a FITS-style extension.

\subsection{Axes and world coordinate systems (WCS)}
\label{sec:wcs}

Axis information was initially specified using an \texttt{AXIS} array
structure with an element for each dimension of the primary data
array. The axis information was specified in an \texttt{ARRAY} type
structure as for data and variance, and allowed axis labels and units
to be specified. For each axis, coordinates were specified for each
pixel in the corresponding dimension of the data array. This allowed
for non-linear axes to be specified and is similar to the
\texttt{-TAB} formalism later adopted in FITS \citep{2006A&A...446..747G}.

The \texttt{AXIS} formalism worked well for spectral coordinates but
it was not suitable for cases where the world coordinate axes are not
parallel to the pixel axes, such as is often the case with Right
Ascension and Declination. Nor did it provide the meta-data needed to
allow the axis values to be transformed into other coordinate systems,
for example when changing a spectral axis from frequency to velocity or
sky axes from ICRS to Galactic. A more general solution was required and this
prompted the development of the AST library
\citep[][\ascl{1404.016}]{1998ASPC..145...41W} with an object-oriented approach to
generalized coordinate frames.

A full description of the data model used by AST is outside the scope of
this paper, but in outline AST uses three basic classes - ``Frame'',
``Mapping'' and ``FrameSet'':

\begin{description}

\item[\texttt{Frame}] - describes a set of related axes that are used to
specify positions within some physical or notional domain, such as ``the
sky'', ``the electro-magnetic spectrum'', ``time'', a ``pixel array'', a
``focal plane''. In general, each such domain can be described using
several different coordinate systems. For instance, positions in the
electro-magnetic spectrum can be described using wavelength, frequency or
various types of velocity; position on the sky can be described using
various types of equatorial, ecliptic or galactic coordinates. A Frame
describes positions within its domain using a specified coordinate
system, but also encapsulates the information needed to determine the
transformation from that coordinate system to any of the other
coordinate systems supported by the domain. It should be noted that a
Frame does not include any information that relates to a different
domain. For instance, a SkyFrame has no concept of ``pixel size''.

Several Frames can be joined together to form a compound Frame describing
a coordinate system of higher dimensionality.

\item[\texttt{Mapping}] -  describes a numerical recipe for transforming
an input vector into an output vector\footnote{Most Mappings also describe the
inverse transformation.}. Most importantly, a Mapping makes
no assumptions about the coordinate system to which the input or output
vector relates. AST contains many different sub-classes of Mapping that
implement different types of mathematical transformations, including all
the spherical projections included in the FITS WCS standard. Several
Mappings may be joined together, either in series or in parallel, to form
a more complex compound Mapping.

\item[\texttt{FrameSet}] - encapsulates a collection of Frames, together
with Mappings that transform positions from one Frame to another. These
are stored in the form of a tree structure in which each node is a Frame
and each link is a Mapping. Within the context of an NDF, the root node
always describes pixel positions in the form of \texttt{GRID} coordinates (see
below). Each other node represents a coordinate system into which pixels
positions may be transformed, and will often include celestial or
spectral coordinate systems. Facilities of the FrameSet class include the
ability to transform given positions between any two nominated Frames, and
to adjust the Mapping between two Frames automatically if either of the
Frames are changed to represent a different coordinate system.

\end{description}

AST support was added to the NDF standard in spring 1998
\citep{2001ASPC..238..129B} by adding a new top-level \texttt{WCS} structure
to NDF. This holds a FrameSet that describes an arbitrary set of coordinate
systems, together with the Mappings that relate them to pixel coordinates.

AST objects were stored simply as an array of strings using their native
ASCII representation, which is more general and flexible
than adopting a FITS-WCS serialization. The NDF library API was modified
to support routines for reading and writing the WCS FrameSet without the
library user knowing how the FrameSet is represented in the hierarchical
model.

The NDF library manages a number of WCS Frames specifically intended to
describe coordinate systems defined by the NDF library itself:

\begin{description}
\item[\texttt{GRID}] - a coordinate system in which the first pixel in the NDF
is centered at coordinate (1,1). This is the same as the ``pixel coordinate
system'' in FITS.
\item[\texttt{PIXEL}] - a coordinate system in which (0,0) corresponds to
the pixel origin of the NDF. The transformation between \texttt{PIXEL}
and \texttt{GRID} is a simple shift of origin.
\item[\texttt{AXIS}] - a coordinate system that corresponds to the AXIS
structures (if any) stored in the NDF.
\item[\texttt{FRACTION}] - a coordinate system in which the NDF spans a
unit box on all pixel axes.
\end{description}

The NDF library always removes the above Frames when storing the
WCS FrameSet within an NDF, and re-creates them using the current state
of the NDF when returning the WCS FrameSet for an NDF. This allows the
user to modify the \texttt{AXIS} and pixel origin information in the
NDF without having also to remember to update the WCS information.

\subsection{History}

The \texttt{HISTORY} structure is used to track processing history and
includes the date, application name, arguments and a narrative
description. The structure was first developed for the
\asterix\ package and adopted directly into the NDF data
model. The history was, by design, not expected to be parseable by
application software; applications such as \surf\
\citep[][\ascl{1403.008}]{1998ASPC..145..216J} did nonetheless use the history to determine whether a
particular application had been run on the data, so that the user could be
informed if a mandatory step in the processing had been missed.

Comparing the history structure of \figref{fig:ndf-structure} with
that shown in \figref{fig:asterix}
shows that the components are identical apart from the addition of
some additional fields to the NDF form (user, host and dataset).
The only other change involved the relatively recent addition to increase in resolution the time
stamp to support milliseconds. This was required as computers have
become faster over the years and many processing steps can occur
within a single second. Provenance handling, \secref{sec:provenance},
uses the history block to disambiguate provenance entries and relies
on the timestamp field.

\subsection{Extensions}
\label{sec:more}

The NDF standard included a special place, named \texttt{MORE}, for local extensions to the
model. This allowed instruments and applications to track additional
information without requiring standardization. The only rule was that
each extension should be given a reserved name (registered informally within the
  small contemporary community) and that the data type
would define the specific data model of an extension.
Some
applications, for example, went so far as to include covariance
information in extensions to overcome limitations in the default error
propagation model for NDF \citep[for example \specdre;][\ascl{1407.003}]{SUN140}.

Three extensions proved so popular that they are now effectively part of the NDF
standard, but for backwards compatibility with existing usage they
cannot be moved out of the extension component. These extensions
covered FITS headers, provenance tracking and data labels and are
described in the following sections.

\subsubsection{FITS headers}
\label{sec:FITSheaders}

FITS headers, consisting of 80-character header cards, are extremely
common.  To simplify interoperability with FITS files and to
minimize structure overhead -- as was found from experience with DST --
it became commonplace to store the header \emph{as is}
as an array of 80-character strings matching the FITS
convention, rather than attempting to parse the contents and expand
into structures.

\subsubsection{Provenance}
\label{sec:provenance}

\begin{figure}
\includegraphics[width=\columnwidth]{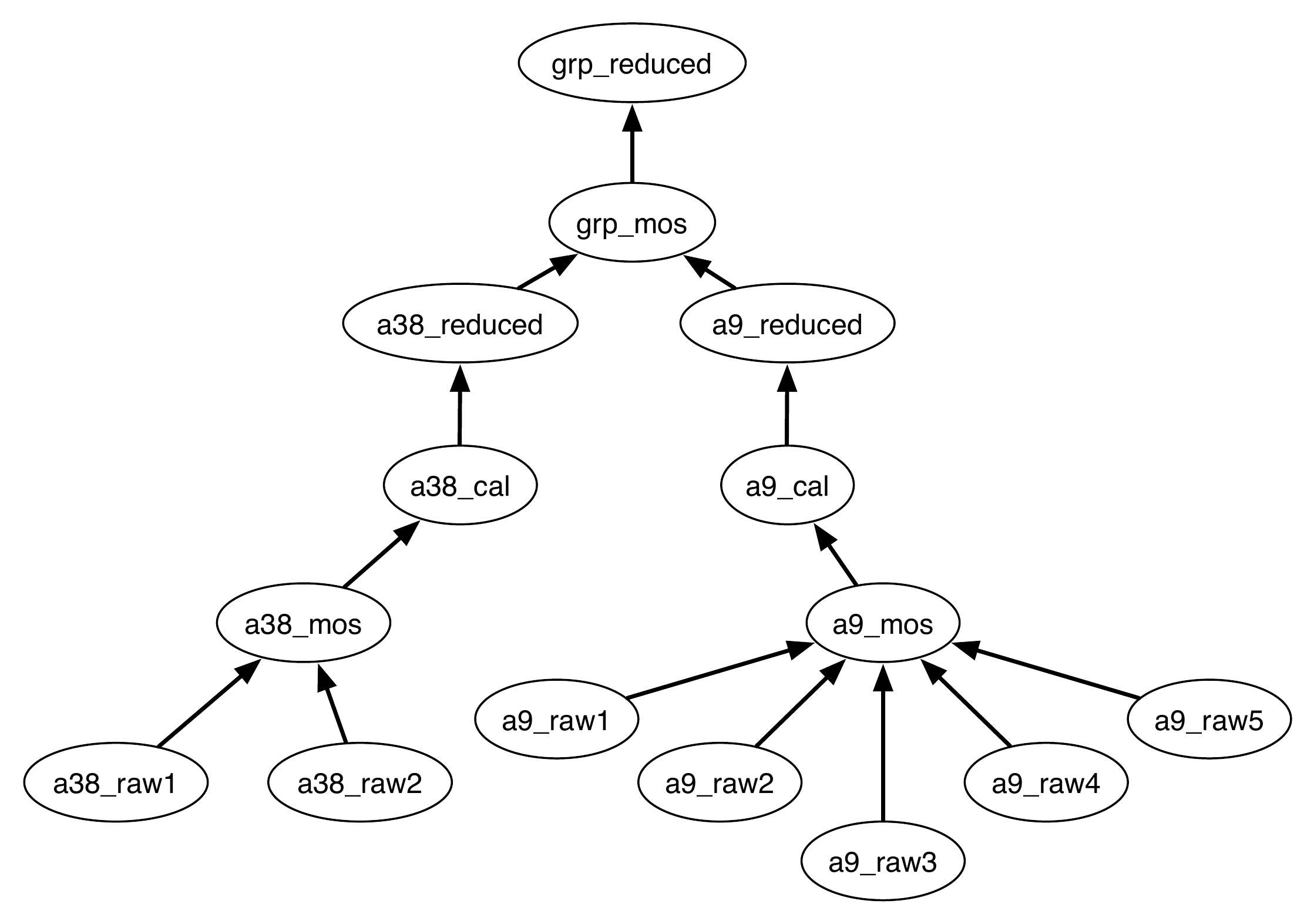}
\caption{A simple provenance tree for an example of two observations
  being reduced independently and then co-added into a final
  mosaic.}
\label{fig:prov}
\end{figure}

In the late 1980s disk space and processing power were more highly
constrained than they are now. At the time, therefore, it
seemed prudent to limit history propagation to a single ``primary''
parent.  As a consequence the NDF library ensures that each application
copies history information from a single primary parent NDF to each
output NDF, appending a new history record to describe the new
application. This means that if many NDFs are combined together by a
network of applications, then each resulting output NDF will, in
general, contain only a subset of the history needed to determine all
the NDFs and applications that were used to create the NDF. It would
of course be possible to gather this information by back-tracking
through all the intermediate NDFs, analyzing the history component of
each one, but this depends on the intermediate NDFs still being
available, which is often not the case.

In 2009, it was decided that the inconvenience of this ``single line of
descent'' approach to history was no longer justified by the savings in
disk space and processing time, and so an alternative system was
provided that enables each NDF to retain full information about all
ancestor NDFs, and the processing that was used to create them \citep{2009ASPC..411..418J,2011tfa..confE..42J}. Thus
each NDF may now contain a full ``family tree'' that goes back as far as
any NDFs that have no recorded parents, or which have been marked
explicitly as ``root'' NDFs. Each node in the tree records the name of
the ancestor NDF, the command that was used to create it, the date and
time at which it was created, and its immediate parent nodes. Each
node also allows arbitrary extra information to be associated with the
ancestor NDF. Care is taken to merge nodes -- possibly inherited from
different input NDFs -- that refer to the same ancestor. A simple
provenance tree is shown in \figref{fig:prov}.

For reasons of backward compatibility, it was decided to retain the
original NDF History mechanism, and add this new ``family tree''
feature as an extra facility named ``Provenance''.  Originally the
family tree was stored as a fully hierarchical structure within an NDF
extension, using raw HDS. However, given the possibility for
exponential growth in the number of ancestors, the cost of navigating
such a complex structure quickly became prohibitive. Therefore,
storage as raw HDS was replaced by an optimized bespoke binary format
packed into an array of integers (the provenance model is outlined in
\ref{app:prov}).

\subsubsection{Quality labels}

Individual bits in the quality mask can be addressed using the
\texttt{BADBITS} attribute, but the NDF standard did not allow for
these bits to be labeled. This confusion was solved first in the
\iras\ package \citep[][\ascl{1406.014}]{SUN165} which added a \texttt{QUALITY\_NAMES}
extension associating names with bits. \KAPPA\ tasks \citep[][\ascl{1403.022}]{SUN95} were then
modified to understand this convention and allow users to enable and
disable masks by name.

\subsection{Library features}

Above and beyond the core data model, the NDF library provides some
additional features that can simplify application development. The
library uses an object-oriented interface, despite being
developed in Fortran, and the python interface, \texttt{pyndf}, provides a full object
interface whereby an NDF object can have methods invoked upon it and
provides direct access to object attributes such as dimensionality and
data label.

\subsubsection{Component propagation}

The NDF library allows output NDFs to be constructed from scratch, but
it also recognizes that many applications build their output dataset(s)
using one of the input datasets, termed the ``primary'' input, as a
template. In this case, much of the primary input data may need to be
propagated to the output unchanged, with the application attending only
to those components affected by the processing it performs. For example,
scaling an image would not affect coordinate data, nor many other
aspects of an NDF. Alternatively, an application might not support
processing of certain NDF components and would then need to suppress
their propagation to avoid creating invalid or inconsistent output.

These requirements are accommodated through use of a component
propagation list that specifies which of the input NDF components should
be propagated directly to the output. A default propagation list is
provided which applications can easily tailor to their specific needs.
This helps to ensure uniformity in application behavior. The default
behavior of the propagation list also encourages the convention that
applications should copy any unrecognized extensions unchanged from
primary input to output. Applications are again free to modify this
behavior, but in general do not do so.

\subsubsection{Data sections}

The NDF library allows a subset of the array to be selected using a
slicing syntax that can work with a variety of coordinate systems. The
user can specify the section in pixel coordinates (where the pixel
origin is taken into account) or in world coordinates, in which case the
supplied WCS values are transformed into pixel coordinates using the
transformations defined either by the legacy \texttt{AXIS} scheme or
the full AST-based WCS scheme, as described in \secref{sec:wcs}.
The section can either be specified using bounds or a center coordinate
and extent. If the section is specified using WCS values, the section of
the pixel array actually used is the smallest box that encloses the
specified WCS limits. Example sections are
\begin{verbatim}
myimage(12h59m49s~4.5m,27.5d:28d29.8m)
\end{verbatim}
to extract an area from an image where the limits along Right
Ascension are specified by a center  and a half width, and the extent
along the Declination axis is defined by bounds;
\begin{verbatim}
myspectrum(400.:600.)
\end{verbatim}
where the spectrum is truncated to a range of, say, 400 to 600\,km/s;
\begin{verbatim}
mycube(12h34m56.7s,-41d52m09s,-100.0:250.0)
\end{verbatim}
where a spectrum is extracted from a cube at a particular location and
also truncated; and
\begin{verbatim}
mycube(55:63,75~11,-100.0:250.0)
\end{verbatim}
where a subset of a cube is extracted where the first two coordinates
specify pixel bounds and the third coordinate is a world coordinate
range. In all these examples a colon indicates a range, and a tilde
indicates an extent, so in the final example \texttt{55:63} means
pixels 55 to 63 inclusive, and \verb|75~11| means 11 pixels centered
on Pixel 75.

\subsubsection{Chunking \& blocking}

Mapping large data arrays into memory can use considerable resources
and astronomical data always seem to be growing at a rate slightly
exceeding the capacity of computers available to the average astronomer. The NDF
library provides an API to allow subsets of a data array to be mapped
in chunks to allow data files to be processed in smaller pieces. The
concept of blocking is also used to allow a subset of the image to be
mapped in chunks of a specified dimension. The distinction between
chunking and blocking is significant in that chunking returns the data
in sections that are contiguous in memory for maximum efficiency
whereas blocking allows spatially related pixel data to be returned, at
the expense of a performance loss in reordering the data array.

\subsubsection{Automated Foreign-format conversion}

In many cases an astronomer would like to use a particular NDF
application on a FITS file without realizing that the application does
not natively support FITS. If foreign-format conversion is enabled, the
NDF library will run the appropriate conversion utility based on the
file suffix and present the temporary, NDF, file to the application. If the
user specifies an output file with a particular foreign-format suffix
the NDF library will then create a temporary NDF for the output and convert
it when the file is closed.

\subsubsection{Automated history}

The library API was designed to make history updating as easy
as possible: by default, if the history structure exists the NDF
library automatically adds a new entry when a file is written
to, and if the history structure is missing no entry is recorded.

\subsubsection{Event triggers}

Callback routines can be registered to be called in response to files
being read from, written to, opened or closed. This facility is used
by the NDG library \citep{SUN2} to enable provenance handling
as a plugin without requiring direct changes to the NDF library itself.

\section{Lessons learned}
\label{sec:lessons}

For a variety of reasons, the Starlink software, and hence NDF,
did not achieve much traction outside the UK and UK-affiliated
observatories. In particular
the US astronomical community that closely adhered to FITS until the
advent of HDF. This is regrettable as NDF proved to be a rich and
flexible data model that has aged well against mounting requirements
from data processing environments. Perhaps the ultimate lesson learnt
is that data formats and models are adopted and retained for sociological reasons
as much as for technical reasons.

\subsection{Key successes}
\label{sec:success}

NDF is a successful data model: it achieves its goal of supporting
broad interoperability of astronomical data between applications in a
pipeline and between pipelines, across multiple wavelengths, and
across time.  It did this without inhibiting applications from
including any and all of the application- or instrument-specific
features they needed to preserve.  Below, we try to tease apart some
separately important strands of this design, but summarize the key
points here.

\subsubsection{Do not try to do everything}

There are two incompatible approaches to designing a data model. One
extreme aspires to think of everything that is needed for
all of astronomy, and to design a model where all metadata are
described and everything has its place. This is the approach taken by
the IVOA \citep[see e.g.][]{2012arXiv1204.3055M}. This is an
intelligible and worthy goal and it is clear that
much can be gained if such data models can be utilized, especially if
related models (spectrum and image when combined into a data cube)
share common ground.  In our experience, however, this approach leads to long and
heated discussions that generate data models that are never quite
perfect and continually need to be tweaked as new instrumentation and
metadata are developed.

Despite much detailed discussion and inventiveness at the time the
model was devised, the NDF model remains modest: \emph{NDF does not
  try to do everything}, but instead adds a bare minimum of useful
structure.\footnote{Einstein did not quite say `everything should be
  as simple as possible, but no simpler', but the thought is no less
  true for being a misquotation, and he might as well get the credit
  for it.}  Astronomy data are fundamentally rather simple, and the
elements in the NDF data model represent both a large fraction of the
information that is necessary for high-level understanding of a
dataset, and a large fraction of what is readily interoperable between
applications without relying on detailed documentation.

For NDF there was a need to generate a usable model quickly that took
concepts that were generically useful, leaving instrumental details to
extensions. These extensions were ignored by generic software packages
although clear rules were made regarding how extensions should be
propagated. This approach allowed for NDF to grow without
requiring the model or the NDF library to understand the
contents of extensions.

This approach of being flexible and not attempting to solve everything
at the first attempt has been very successful, not only in enabling new
features to be added to NDF once the need became obvious, but also
in so far that wavelength regimes not initially involved in the
discussion can make use of NDF without requiring that the core data
model be changed.

\subsubsection{Hierarchy can be useful}

Initially it was very hard to convince people that hierarchical data
structures were at all useful. This can be seen in the Wright-Giddings
layout of a simple astronomical image file using HDS. Over the years
the adoption of hierarchy has been an important part of the NDF
experience, and application programmers quickly began to realize the power of
being able to relocate or modify parts of a data file,
to change the organization, or to copy related sections to a new file.
NDF extensions may contain other NDF structures, allowing
software packages to extract those NDFs, or simply focus on them,
without regard to the enclosing data module; this becomes very intuitive and
obvious once it is learned. A flat layout would require additional
grouping metadata to understand which components were related, but this
is neatly handled by a simple hierarchy.

\subsubsection{Allow `private' data (namespaces)}

The NDF model is strongly \emph{hierarchical}.  This both makes it
easy to find and manipulate data elements, and makes it easy for
applications to \emph{extend the model}, by providing a space for
private data, which does not interfere with the shareability or
intelligibility of the dataset as a whole.

The NDF model, by design, supports only a subset of the information
associated with a realistic dataset; it trivially follows that most
data providers will want to include data that are not described within
the NDF model, and if the model could not accommodate these extra
data, it would be doomed to be no more than a `quick-look' format.

As described above (\secref{sec:more}), the NDF model includes a
\texttt{MORE} tree, where particular applications can store
application-specific data using any of the available HDS structures
(\secref{sec:hds}).  Some applications treated this as a black box;
others documented what they did here, making it an extended part of
the application interface, so that their `private' data were effectively
public, and were largely standardized in practice without the long
drawn-out involvement of a formal standards process.  Thus, different
elements of the data within an NDF file were painlessly standardized
by different entities, on different timescales and in response to
direct community demands.

When applications wished to take advantage of the possibilities here,
they registered a label within the \texttt{MORE} tree, and were deemed
to `own' everything below that part of the tree.  The community of NDF
application authors was small enough that this could be done
informally, but if this were being designed now, then a simple
namespacing mechanism (using perhaps a reversed-domain-name system)
would be obvious good practice.

\subsubsection{Standardized features aid application writers}

Once application writers understand that there is a standard place for
error information, quality masks and other features of the standard
data model, they begin to write application software that can use
these features. Not having to use a heuristic to determine whether a
particular data array represents an image or an error can allow the
application writer to focus more on the algorithms that matter. Once
users of the software understand that errors and masks are an option
they begin to have an expectation that all the software packages will
handle them. This then motivates the developer to support these
features, creating a virtuous circle which tended to improve all of
the software in the NDF community.
This is especially true if the core concepts of the data
model are simple enough that the learning curve is small.

\subsection{Other good features}

In addition to the key successes there were a number of
other lessons learned while using the data model over the years.

\subsubsection{Round-tripping to other data formats}
\label{sec:round_trip}

For a new format to be used it must be possible to convert to and from
existing formats in a safe and reliable manner without losing
information \citep[see for example,][for a discussion of this problem with FITS and HDF files]{1995ASPC...77..229J}.
The NDF model on HDS format files has always been one format amongst many
formats in astronomy and much work has been expended on providing
facilities to convert NDF files into FITS and IRAF format, and
\emph{vice versa} \citep{SUN55,1997STARB..19...14C}.  In particular,
special code was used to recognize specific data models present in
FITS files to enable a more accurate conversion of scientific
information into the NDF form. Support was also added for the
\emph{multispec} data model in IRAF \citep{1993ASPC...52..467V} to ensure
that wavelength scales were not lost. A description of how NDF maps to
FITS is given in \ref{app:ndf2fits}.

\subsubsection{Adoption of FITS header}

Given hierarchical structures the default assumption might be to use
arrays of keyword structures where each structure would contain the
value, the comment and the unit. Alternatively, one might simply drop the unit
and comment and just use keyword/value pairs. These approaches turned
out to be extremely slow and space inefficient so the project took the  pragmatic approach
of standardizing on an array of characters formatted
identically to a FITS header.
This allowed the FITS to NDF conversion
software to ignore new FITS header conventions as they were developed,
deferring this to the application writer who can decide whether a
particular convention is important. Without this pragmatic approach
the user would have to re-convert the files whenever a new convention
is supported and the NDF data model would have to be extended to
support such features as hierarchical keywords and header card units.
In practice the AST
library \citep{1998ASPC..145...41W} is often used for processing the FITS
header from an NDF so this causes few problems and simplifies the NDF
organization, allowing an application to understand new FITS header
conventions simply by upgrading the AST library.

\subsubsection{Duck typing}

\begin{figure}[t]
\begin{center}
\includegraphics[width=\columnwidth]{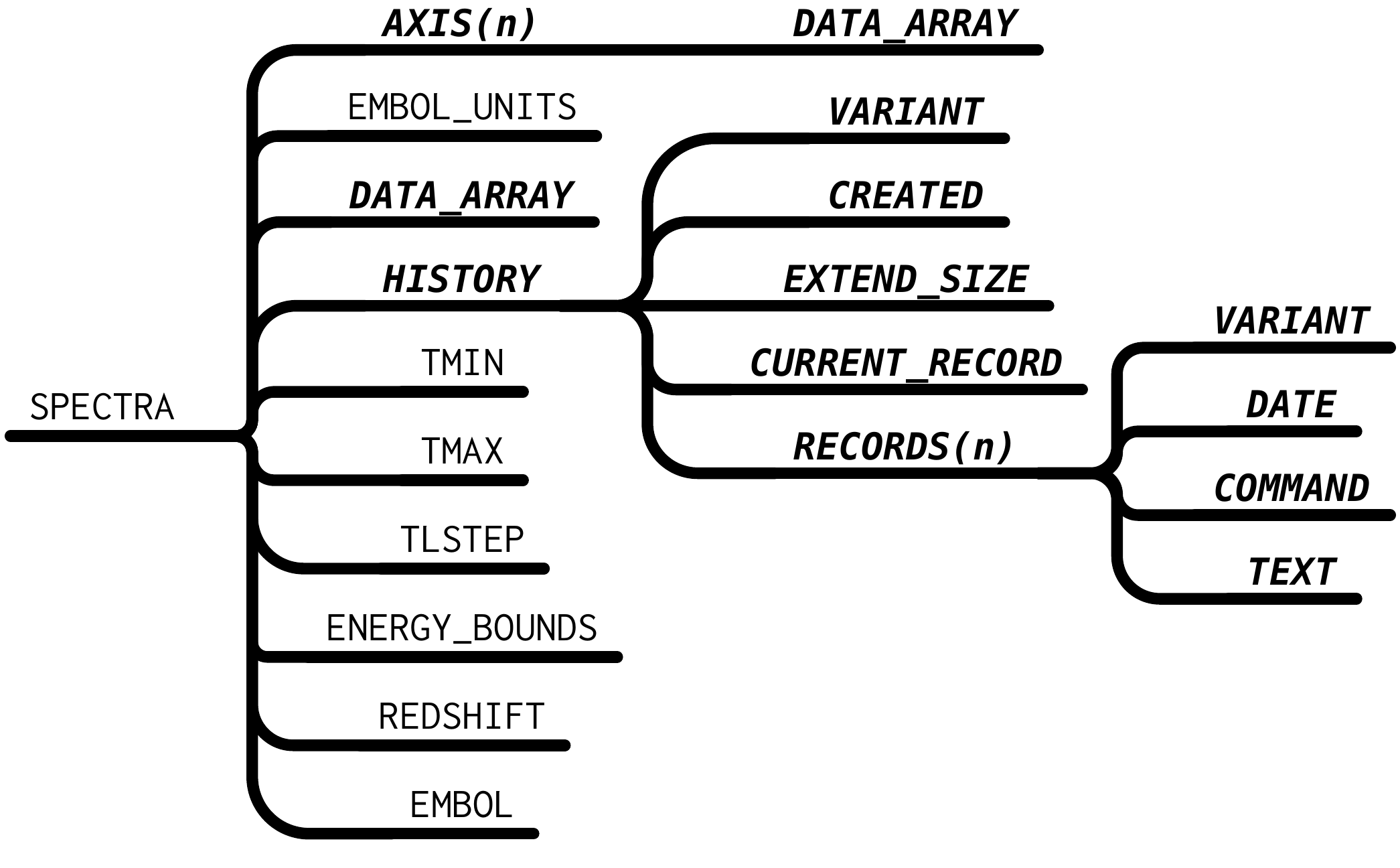}
\end{center}
\caption{Partial layout of the structure of an \asterix\ HDS
  \texttt{spectra} cube file made in 1992. This file, written after the NDF
  standard was written, adopted some of the elements of the
  NDF model and can be processed by NDF-aware applications despite not
  conforming completely to the standard. Components recognized by NDF
  are labeled in italics}
\label{fig:duck}
\end{figure}

In the early years of NDF there were many files in existence that did
not quite conform to the standard and new files were being created by
existing applications that did not use the main NDF library but
nonetheless modified their structures to emulate an NDF (see for
example \figref{fig:duck}). The core NDF library thereby took an
inclusive approach to finding NDF structures, with the only requirement
being the existence of a \texttt{DATA\_ARRAY} item. If other compliant
structures were found, the NDF library would use them and ignore items
that it didn't understand. This ``duck typing''\footnote{See for
  example \url{http://en.wikipedia.org/wiki/Duck_typing}} approach
proved to be very useful and eased adoption of NDF. The NDF
library uses this feature to scan for NDF structures within an NDF or
an HDS container. For example, the \gaia\ visualization tool \citep[][\ascl{1403.024}]{2009ASPC..411..575D}
will scan an HDS file for all such NDF structures and make them
available for display. The data type of the structure -- which may be
thought of as the class name -- is not used by the library when
determining whether a structure is an NDF.

\subsubsection{Extensible model}

The initial NDF design document did not profess to know the future and
was deliberately designed to allow new features to be added as they
became necessary. The adoption of a character array matching an
80-character FITS header was an early change but there have also been
changes to support world coordinate objects
\citep{2001ASPC..238..129B}, data compression algorithms
\citep{2008ASPC..394..650C} and provenance tracking
\citep{2009ASPC..411..418J}. The NDF standard therefore has a proven
ability to continue to evolve
to meet the needs of modern astronomy data processing.

NDF combined with HDS was a conscious decision to not develop an
interchange or archive format. HDS was designed solely to be accessed
through a reference implementation library and API and many of the NDF
features were directly integrated into the reference NDF library without being
specified in the core data model.  This yielded great flexibility
regarding exactly how the data are stored and greatly simplified
future enhancements (and general future-proofing) because changes to
the data format and how data are accessed could all be hidden from
applications by changes made in one place. However, care needs to be taken
when rolling out updates: software linked with old library versions
may not be able to read newer data. Indeed, HDS includes an internal
version number and can read earlier versions of the data format, but if a
newer version is encountered than supported by the library HDS will
know about it and report the problem.

\subsection{Extension Difficulties}

Some aspects involving the use of NDF caused confusion amongst users,
particularly when extensions were involved.

\subsubsection{Extension complexity}

\citet{SGP38} provided an algorithm for designing simple elemental
building blocks that could be reused. However, many extensions created
in the early years of NDF became more complex than was strictly
necessary by ignoring the design decisions of NDF and using highly
complex HDS structures.  The lack of a standard table structure was
only partly to blame.  In many cases it would have been beneficial to
use NDF structures in the extensions, which would have allowed the NDF
library to read the data (by giving the full path to the structure)
without resorting to low-level HDS API calls.  The NDF structures
would also have been visible to general-purpose tools for
visualization. Future designers should not expect their respective
community to design any elemental structures or the API to support
them.

\subsubsection{Name versus type}
\label{sec:name_v_type}

Extension implementors often ignored or misunderstood another
fundamental precept of standard structures, the difference between the
name and type of a structure.  They relied on the structure name,
often omitting to define a type.  The data type defines the content,
semantics, and processing rules.  The name is just an instance of the
particular structure.  This approach could lead to name clashes, or
restrict to only one instance within a given structure.  Software
other than the originator's encountering the structure would not know
the meanings and how to handle the structure's contents.  Note that
this happened at a time when object-oriented concepts were not widely
known.  Again more oversight would have helped.

The same confusion is evident in FITS extensions where there is no
standard header to record the extension's type,\footnote{This type is
not to be confused with the extension type (\texttt{XTENSION}) which
defines how the data are stored rather than what they represent.}
perhaps borne of FITS's lack of semantics, although Wells had added
the extension keywords partly with Starlink hierarchical data in mind.
Keyword \texttt{EXTNAME} is often used to attribute meaning.
Reserving \texttt{EXTNAME} for the NDF component path, and adding an
\texttt{EXTTYPE} header permitted a round trip between NDF and FITS
(see \secref{sec:round_trip}).  \texttt{EXTTYPE}, proposed by
\citet{1993EXTKEYWORDS}, was never adopted into the FITS standard
\citep{1993OGIPMIN}.

\section{Areas for improvement}
\label{sec:improve}

As the NDF data model has been used we have come to realize that some
parts of the standard could do with adjustment and other enhancements
should be made. The following items could be implemented and the
current standard does not preclude them. One caveat is that there are
no longer any full-time developers tasked with improving NDF and any
improvements would be driven by external priorities of the key
stakeholders (as was the case with the recent addition of provenance
and data compression).

\subsection{Quality masking}

The initial design for the quality mask used a single unsigned byte
allowing eight different quality assignments in a single data file. The
design did not allow for the possibility of supporting larger unsigned
integer data types. This restriction should be raised to allow more
assignments. The SMURF map-maker
\citep[][\ascl{1310.007}]{2013MNRAS.430.2545C} already makes use of more
than eight internally and uses an unsigned short. When the results are
written out mask types have to be combined if more than eight were present
in the output data file.

\subsection{Table support}

Tables were added to FITS \citep{1988A&AS...73..365H} during the
initial development of NDF and the need for tables was considered a
lower priority in the
drive for a standardized image model and table support was never added
to NDF.  The omission of a table model was belatedly addressed in the
early 1990s from outside Starlink.  Giaretta proposed and demonstrated
that HDS could be used to store tables, being especially suited to
column-oriented operations.  Such a \texttt{TABLE} data structure was
later interfaced through the CHI subroutine library \citep{SUN119} for
the \catpac\ package \citep{SUN120}.  This initiative failed, however,
due to a lack of documentation and promotion.  The effort was not
completely wasted.  The \fitstondf\ conversion tool \citep{SUN55}
creates an extended form of the \texttt{TABLE} model (see
\ref{app:hdstable}) to encapsulate within an NDF the ancillary FITS
tables associated with image data.

Starlink software eventually took the pragmatic solution of using FITS
binary tables \citep{1995A&AS..113..159C} for output
tables.\footnote{Text formats such as tab-separated lists and
Starlink's own Small Text List format supported by the CURSA package
\citep{2001ASPC..232..314D} were also more popular than HDS tables.}
This cannot solve the problem of integrating data tables into image
data files and the format would benefit from a native data type. The
JCMT raw data format instead uses individual 1-dimensional data arrays
to store time-series data but this is inefficient and adds programming
overhead.

\subsection{Flexible variance definitions}

The adoption of variance as a standard part of NDF was an important
motivator for application writers to add support for error propagation and
almost all the Starlink applications now support variance. The next
step is to support different types of errors, including
covariance \citep[see e.g.][]{1992ESOC...41...47M}. This has been proposed many times \citep[see
e.g.][]{1991STARB...8...19M} but is a very difficult problem to solve
in the general case and may involve having to support pluggable
algorithms for handling special types of error propagation.

\subsection{Data checksums}

The FITS \texttt{DATASUM} facility \citep{2012arXiv1201.1345S} is very useful and NDF should support
it. Ideally it should be possible to generate a reference checksum for
a structure. It may be that this has to be done in conjunction with
HDS.

\subsection{Character encodings}

The \texttt{\_CHAR} data type in HDS uses an 8-bit character type,
assumed to contain only ASCII. It was designed long before Unicode came to exist
and has no support for accented characters or non-ASCII character
sets. Multi-byte Unicode should be supported in any modern format to
allow metadata to be represented properly. HDS cannot support the
storage of common astronomical unit symbols such as $\upmu$m or \AA.
It may be possible to use the automatic type conversion concept
already in use for numeric data types
to allow Unicode to be added to NDF without forcing every application
to be rewritten. If an application uses the standard API for reading character
components they will get ASCII even if that involves replacing Unicode
characters with either normalized versions of characters outside the
ASCII character set (for example, dropping accents) or whitespace. A new API
would be provided for reading and writing Unicode strings.

\subsection{Provenance growth}

Whilst the provenance handling works extremely well and is very useful
for tracking what processing has gone into making a data product, the
provenance information can become extremely large during long and
complicated pipeline processing such as those found in products from
ORAC-DR \citep[][\ascl{1310.001}]{2008AN....329..295C,2015oracdr}.  There can be many thousands of
provenance entries including some loops where products are fed back
into earlier stages because of iterations.  The provenance tracking
eventually begins to take up a non-trivial amount of time to collate
and copy from input files to output files. One solution to this may be
to offload the provenance handling to a database during processing,
only writing the information to the file when processing is complete
and the file is to be exported. It would be fairly straightforward to
modify the NDG library to use the Core Provenance Library
\citep{Macko:2012:GPL:2342875.2342881}.

\subsection{Library limitations}

Whilst there are many advantages to having a single library
providing the access layer to an NDF, there is also a related problem
of limitations in this one library causing limitations to all
users. In particular the current NDF library is single-threaded due to
use of a single block of memory tracking access status. Furthermore
the HDS library itself is also single-threaded with its own internal
state. As more and more programs become multi-threaded to make use of
increased numbers of cores in modern CPUs this limitation becomes more
and more frustrating.

Another issue associated with the library is the use of 32-bit
integers as counters in data arrays. It is now easy to imagine data
cubes that exceed this many pixels and a new API may be needed to ease
the transition to 64-bit counters.

Whilst the HDS library is written entirely in ANSI C, the NDF and
related ARY \citep{SUN11} and NDG \citep{SUN2} libraries are written
mainly in Fortran.
This puts off people outside the Starlink community and
adds complications when providing interfaces to higher-level languages
such as Python, Perl and Java. Indeed a subset of the NDF model  was
written as a Java layer on top of the C HDS library, precisely to
avoid the added dependency to the Fortran runtime library.

Ideally NDF would be rewritten in C and be made thread-safe.

\section{Social, Political and Economic Considerations}
\label{sec:social}

We draw attention in this paper to many technical aspects of the NDF
design that remain relevant today. However, the data format and
associated model has its
roots in the social, political, economic and, indeed, technological
circumstances of the late 1980s. So it is relevant to examine what
effect these had, how they have changed, and how they affect data
format choices being made today. We also discuss briefly some of the
political and economic issues that data format developers need to
address when promoting their products.

\subsection{Historical Perspective}

In the period when the NDF was being designed, the Starlink Project
was funded to support the computing needs of astronomical research in
the UK, across all wavebands. On the software side, its remit was to
centrally provide, curate and distribute software, to foster
collaboration between UK software development projects, and to develop
standards for their use. Together with central purchasing of computer
hardware, good networking and a centrally managed team of system
administrators located around the UK, the project was an innovation
that attracted much interest.

Some, however, saw an element of socialism in this arrangement and
argued that, in a more free-market approach, individual research
groups should develop software (and by implication data formats and models)
independently and that the fittest should survive and be used by other
groups. That, indeed, was the default situation in most countries. In
the UK, however, two key objections trumped this argument. One was
that the quality, reliability, maintainability and suitability for
long-term re-use of software and standards produced in this way was
inadequate, essentially because those embarking on such projects,
while being talented scientists, at that time typically lacked software engineering
skills. The second was that resources for astronomy are limited and
that funding multiple similar projects in order to later discard most
of the work in favor of just one is wasteful. Consequently, UK-wide
collaborative development of software and standards, including data
formats, was the option that received funding.

The broad range of astronomy to be supported by Starlink and the
consequent diverse data requirements was a severe challenge and
exposed the Project to many issues that other groups had to face only
later on. As we have described, the problem was made more tractable by
using a single very flexible and extensible data format, rather than a
set of individual ad-hoc solutions for each branch of astronomy. So
the motivation for developing NDF was, in essence,
economic. It was cheaper to take that route.

Starlink's success was sometimes judged by international
take-up\footnote{At one time overseas users accounted for about
 12\,\% of the user population for the Starlink Software Collection
 as a whole \citep{1992STARB..10...30L}.} of its
software and standards, but its funding was primarily in support of UK
astronomy, so software promotion and support activities outside the UK
only took place on a best-efforts basis. While widespread adoption of
NDF outside of UK projects would have been a bonus, it was
not a funded goal and, indeed, the Project took few steps even to
monitor it.

A key consequence of this ``funding boundary'' was that Starlink staff
were not involved in discussions about data handling for new
projects unless there was direct UK involvement. Such early
discussions are typically crucial in securing ``buy-in'' from potential
new users of software. They allow the software's capabilities and the
costs and benefits of using it to be explored, and provide a timely
opportunity to add features that may be of special interest to that
project. By restricting this activity to UK projects, usage of
Starlink's NDF inevitably became much more widespread
within the UK and its overseas observatories than elsewhere.

\subsection{Current Environment}

Nevertheless, widespread international adoption will be uppermost in
the mind of anyone contemplating a new data format today. This paper's
authors have been reflecting on the relevant issues over many years,
but unfortunately we cannot offer a simple winning formula. Even the
most seasoned of standards organisations will struggle to reconcile
the vested interests of particular groups with the wider interests of
their community and to inspire the confidence needed for widespread
adoption of a standard. Perhaps the best that can be said is that
designing a standard requires a great deal of discussion and
consultation. This can be lengthy and needs robust processes and good
leadership, but it is something that has to be done and should not be
rushed. Time spent getting it right is repaid later by extending the
lifetime and uptake of the standard when in service.

While we have no silver bullet, it is perhaps relevant to ask how we
might start out again now, if embarking on a new data format project
with widespread international adoption in today's environment as a
goal. It turns out that many of the problems we faced in the 1980s
have now been addressed by developments in other areas and that if we
seek out current best practice in software development the number of
difficult choices is fairly small.

Despite its successes, it is unlikely that we would take the FITS
route and define a file format without also implementing accompanying
software. This is because it requires the software to be written
independently, which means (inevitably) by different groups - and it
all needs to inter-operate. Specifying not just the bit-patterns but
also the semantic interpretation of data in sufficient detail for
widespread inter-operability is extremely difficult. If one examines,
say, the HTML standards \citep{w3chtml5}, they are very
lengthy and require a huge effort to produce, yet still
inter-operability often fails in practice. Astronomical data formats
have the potential to become much more complex than this.

One must also avoid designing a data format that cannot easily be
implemented in software. This may be because the algorithmic details
are not apparent but turn out to be complex or ambiguous, or because
of the ubiquitous tendency to over-design and to be wildly
over-optimistic about the resources needed for implementation.

For these reasons we would probably consider developing any new data
formats alongside a reference software implementation. This is a
procedure used by most participants in international software
standards discussions in order to guide and inform their negotiations
and to ensure that they have a working product as soon as the standard
is complete. It ensures that the standard and the implementation never
diverge too far and that the data semantics and their implications are
fully understood (these details being difficult to capture other than
in software). In an astronomy context it should be unnecessary for
each participant to develop their own separate software. Rather, a
single collaboratively developed software project might provide a
suitable nucleus for the whole enterprise.

When developing NDF, the discussion surrounding its design
was open, inclusive and collaborative, but the development of software
to implement it occurred later and (despite eventually being released
under an open-source licence) followed what would now be recognized as
``closed source'' development practices. We did not then have the
benefit of modern collaborative software development tools, but the
success of the free open-source software (FOSS) movement has shown the
power of the open-source development model and some of its features
would certainly have been very attractive.

There are, of course, many open-source software tools, environments
and development processes in use in different FOSS projects. However,
the relevant features found in most cases are: i) frequent builds, so
that a functioning development version of the software is always
available, ii) anyone can participate in discussions and iii) anyone
can contribute code (possibly subject to review by others). Such a
system would seem to meet many of the requirements for developing a
data format specification alongside a reference software
implementation while simultaneously stimulating informed discussion,
feedback and, of course, attracting coding contributions which are
always the scarcest commodity in such an undertaking. One should
perhaps add that good management would also be essential and that
fundamental choices such as implementation language and software
dependencies can be critically important.

\subsection{Thoughts on the data format business}

A crucial issue for a project developing or maintaining an
astronomical data format is to ensure that it can continually gain new
users, something that tends to become easier as the project grows and
becomes more capable and better-known. This principally means that it
must work to be adopted by new instrumentation groups that will be
producing important astronomical data in future. This, in turn, means
winning the arguments for and against adopting a particular data
format within these groups.

Although general-purpose application software suites are also big
players in the data format game, new ones start up relatively
infrequently and may take decades to mature. Their capabilities also
tend to lag behind the demands of new instrumentation. They therefore
present less of an opportunity for promoting a new data format. Having
a major software suite on board is definitely worthwhile, but an
existing well-entrenched data format in a major application suite
presents a huge barrier to entry that can only really be addressed by
implementing a transparent compatibility layer (and/or automated
conversion) so that changes to existing applications are
unnecessary. This is more of a technical challenge than a political
one and we will asume that such a layer will always be put in place,
so that access to existing applications is not an issue in deciding to
adopt a data format.

The problems that then arise in discussions with instrumentation
groups vary, but often economic factors enter. Such a group is most
likely to adopt a format that already closely matches their
requirements, so that they minimise their work. But this only applies
if the software is open, in the sense that they can make any changes
themselves. If changes need to be implemented (or integrated) by a
different data format project, it can be a serious disincentive unless
the data format project commits to making the changes well in advance
and enjoys a significant level of trust. With an open-source data
format development project, however, the openness is built-in and this
barrier should be much reduced. Nevertheless, any changes would still
need to avoid regression in the software that already
exists.\footnote{The alternative is to fork the project, but there
  will be reluctance to do this because support for the code that
  already exists will be lost.}  This, in itself, can be a serious
obstacle if the system is already complex, as most data format
software is. If the data format is designed to be extensible, however,
this burden is greatly reduced because extending the system consists
largely of making additions and not changing what already exists.

Different problems can arise if a new instrumentation group is large
and/or particularly well-funded. Such groups may have little incentive
to collaborate for economic reasons and might be tempted to start from
scratch themselves producing, in effect, a rival project. This is
arguably how most data formats arise in the first
place. Unfortunately, the result may have limited applicability
outside the originating project. Political arguments may sway the
decision and these will be easier to make if the proposed data format
already enjoys wide use. The argument becomes much easier to win,
however, if the data format offering is intrinsically extensible,
because this pretty much guarantees that the work involved in building
on what exists will be substantially less than starting from
scratch. Moreover, it need not involve any compromise to the
instrumentation group's goals because there is no constraint on what
can be added.

If major new instrumentation projects could be harnessed to contribute
to a collaborative open-source data format project with worldwide
adoption as a goal, then major benefits might accrue. Perhaps the most
significant would be that instead of new incompatible formats
regularly arising they could instead be new compatible additions to a
single format of increasing capability. There are many examples where
well-funded initiatives build on an existing FOSS project for their
own purposes and the new developments are then fed back into the main
trunk for all to benefit. There seems to be little reason why
astronomical data formats couldn't develop in a similar, evolutionary
way if openness and extensibility are fully embraced.

\section{Conclusions}
\label{sec:conclusion}

The NDF data model did bring order to the chaos of arbitrary
hierarchical structures and succeeded in the promise of providing a
base specification that can be adopted by many applications processing
data from disparate instrumentation.  The shift from arbitrary use of
hierarchical structures to a data model enforced by a library and API
was extremely important and allowed application developers to know
what to expect in data files.

From its beginnings in the mid-1980s the NDF data model has been used
throughout the Starlink software collection within diverse
applications such as \smurf\ \citep{2013MNRAS.430.2545C}, \ccdpack\
\citep[][\ascl{1403.021}]{1993ESOC...47...39W,SUN139}, \gaia\ and
\KAPPA.

NDF was also adopted by UK-operated observatories.  \figaro\ had a
strong influence on the infrared spectroscopy community in the United
Kingdom and the United Kingdom Infrared Telescope (UKIRT) initially
adopted DST (\secref{app:figaro}) for CGS3 and CGS4
\citep{1993SPIE.1946..547W}. NDF was adopted at UKIRT in 1995 although
a unified UKIRT NDF-based data model for all instruments, involving
HDS containers of NDF structures to handle multiple exposures, was not
adopted until the release of the ORAC system
\citep{2000SPIE.4009..227B}. The James Clerk Maxwell Telescope (JCMT)
initially used a proposed submillimeter standard format known as the
Global Section Datafile \citep[GSD;][formerly General Single Dish
Data]{sun229}. In 1996 SCUBA \citep{1999MNRAS.303..659H} was delivered
using NDF, and a unified NDF raw data model was adopted for ACSIS
\citep{2009MNRAS.399.1026B} from 2006 and SCUBA-2
\citep{2013MNRAS.430.2513H} from 2009. NDF data files are available
from the UKIRT and JCMT archives at the Canadian Astronomy Data Centre
\citep{2008ASPC..394..450E,P01_adassxxiii,2014Economou} and both
telescopes continue to write data using NDF.  The
Anglo-Australian Observatory (AAO) adopted HDS and initially used
DST for instruments such as UCLES
\citep{1990SPIE.1235..562D}. IRIS \citep{1993PASAu..10..298A} could
use both DST and NDF, whereas 2DF \citep{2002MNRAS.333..279L}
used NDF.

The NDF data model also supported a number of data-structuring
experiments.
The model allowed applications written in Fortran to
adopt object-oriented methodologies by adopting NDF as a backing store
and using the self-describing features to represent objects
\citep{1993ASPC...52..199B}.
The HDX framework \citep{2003ASPC..295..221G} was developed around 2002 as a flexible
way of layering high-level data structures, presented as a virtual XML
DOM, atop otherwise unstructured external data stores.  This was in
turn used to develop Starlink's NDX framework,\label{sec:ndx} which allowed FITS
files to be viewed and manipulated using the concepts from NDF.
The NDX experiment was an attempt to directly apply the
lessons of the long NDF/HDS experience -- namely that a small amount of
structure, overlaid on conceptually separate bit buckets, can very
promptly bring order out of chaos.
The experiment successfully demonstrated the viability and power of
the approach, and was used in some Starlink Java applications
(including \treeview\ \citep{2003ASPC..295..445B} and \splat\
\citep[][\ascl{1402.007}]{2005ASPC..347...22D,2014Skoda}); however it
lost out to the more mainstream approach adopted by others
within the VO, and was not more widely adopted.

For the future we are considering the possibility of replacing the HDS
layer with a more widely used hierarchical data format such as HDF5
\citep{Folk:2011:OHT:1966895.1966900}. This would have the advantage
of making NDF available to a much larger community, albeit with NDF
still being in Fortran, and also remove the
need to support HDS in the longer term. NDF and all the existing
Starlink applications would continue to work so long as a conversion
program is made available to convert HDS structures to HDF5 structures.
This would have the added advantage of making it straightforward to
add support for tables natively to the NDF data model. Many of the
concepts in NDF map directly to HDF5. One remaining issue is that HDF5
does not support the notion of arrays of groups so the
\texttt{HISTORY} and \texttt{AXIS} structures in NDF would need to be
remapped into a flatter layout, maybe with numbered components in an
\texttt{AXIS} or \texttt{HISTORY} group.

\section{Acknowledgments}

This research has made use of NASA's Astrophysics Data System.
The Starlink software is currently maintained by the Joint Astronomy
Centre, Hawaii. We thank Jim Peden and Trevor Ponman for providing
comments on the manuscript regarding the early days of HDS and the
development of \asterix. We also thank to the two anonymous referees
for their useful comments.

The source code for the NDF library and the Starlink software
(\ascl{1110.012}) is open-source and is available on Github at
\htmladdnormallink{https://github.com/Starlink}.


\appendix

\section{Early HDS-based data models}
\label{app:chaos}

This section provides an overview of the HDS-based data models
developed within the Starlink ecosystem that influenced the
development of NDF.

\subsection{Wright-Giddings \emph{IMAGE}}
\label{app:image}

\begin{figure}[t]
\begin{center}
\includegraphics[width=0.5\columnwidth]{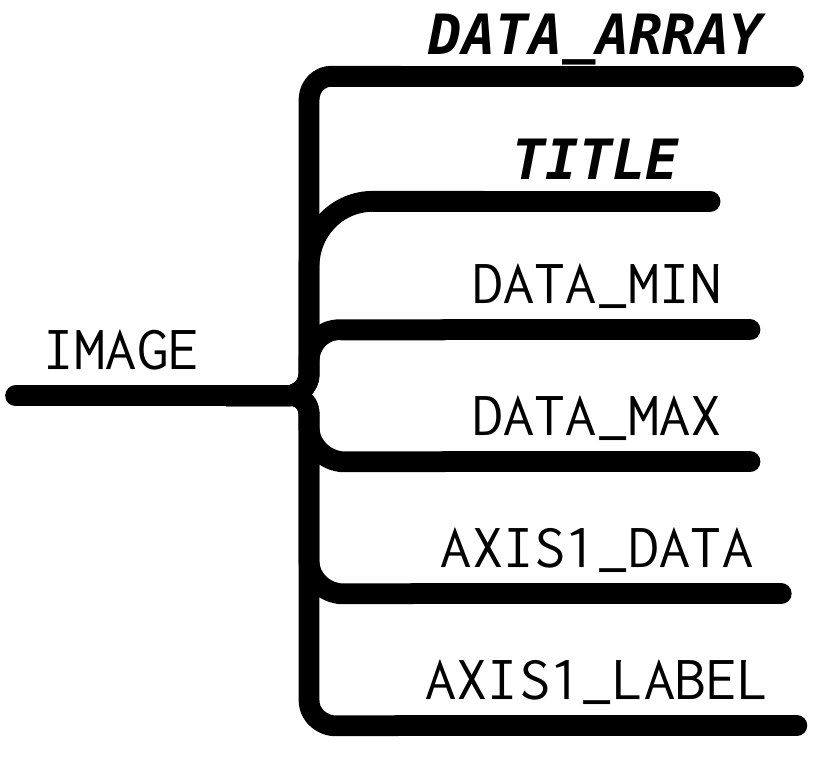}
\end{center}
\caption{Example components of a Wright-Giddings \emph{IMAGE}
  file.  Italicized text indicates the components that are shared with NDF.}
\label{fig:image}
\end{figure}

An early proposal \citep[][but see also \citet{SGP38}]{WrightGiddings1983} introduced the
\emph{IMAGE} organizational scheme. This Wright-Giddings design specified that
data should go into a \texttt{DATA\_ARRAY} item and there should also be
items for pre-computed data minimum and maximum, as well as a value for
an array-specific blank value (similar to the FITS \texttt{BLANK}
header keyword). Errors were represented as standard
deviations and stored in \texttt{DATA\_ERROR} and bad-pixel masks were
stored in \texttt{DATA\_QUALITY}. An example layout is given in \figref{fig:image}.

Prior to HDS becoming generally available, the Starlink Project adopted
the Bulk Data Frame \citep[BDF;][]{1980SPIE..264...70P,SUN4} as part
of its \emph{INTERIM} software environment.  BDF was heavily influenced
by FITS and used many of the same conventions.  Software was provided
to convert BDF format files to HDS using the \emph{IMAGE} model
\citep{SUN96}, and
the \emph{IMAGE} model became reasonably popular, because of its
simplicity, and because of the many BDF files that existed at the time.
There were however
a number of shortcomings with the \emph{IMAGE} design, not the least of which was
that it did not make use of hierarchical structures. The design was
flat and heavily influenced by FITS and BDF.

\subsection{Figaro DST}
\label{app:figaro}

\begin{figure}[t]
\begin{center}
\includegraphics[width=0.6\columnwidth]{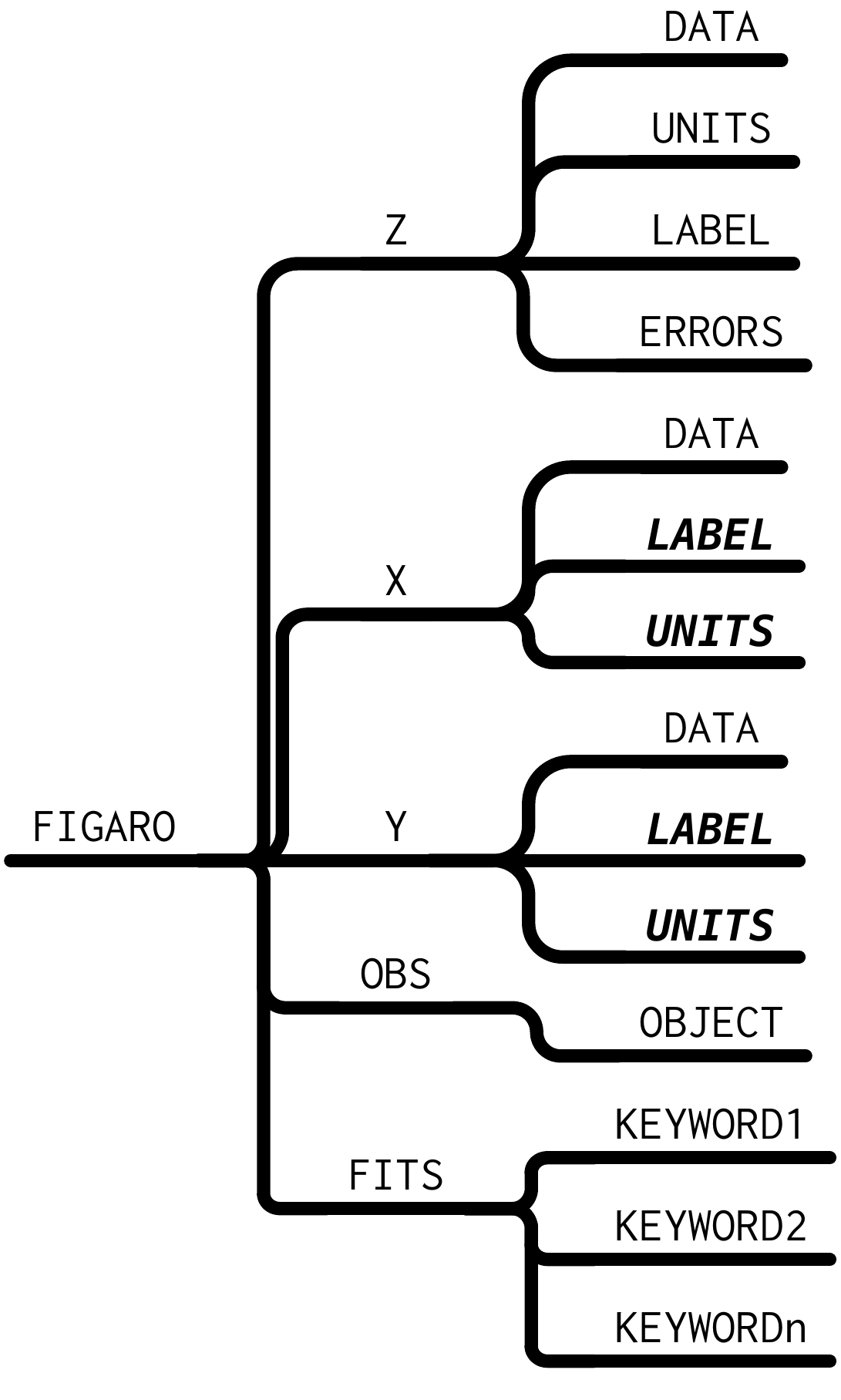}
\end{center}
\caption{Partial model representing the structure of a DST file. Structures make
  use of a hierarchy and reuse concepts in the data array and axis
  definition.  Italicized text indicates the components that are shared with NDF.}
\label{fig:dst}
\end{figure}

The \figaro\ data reduction package
\citep[][\ascl{1203.013}]{1988igbo.conf..448C,1993ASPC...52..219S}
independently adopted a hierarchical design based on HDS. This DST
data model\footnote{The reason for the name has been lost in the mists of
  time but our best guess is that it stood for \emph{\textbf{D}ata
    \textbf{ST}ructures}.} made good use of structures and supported
standard deviations for errors. Axis information was stored in
structures labeled X and Y, and the main image/spectral data were
stored in a structure labeled Z. The main data array was
\texttt{Z.DATA}\footnote{where the `\texttt{.}' indicates a
  parent-child relationship analogous to a directory separator in the
  file system.}. FITS-style keyword/value pairs were encoded
explicitly in a structure called \texttt{FITS} but using scalar
components for each header item. Any comments associated with the FITS
keywords were held in a similar structure labelled
\texttt{COMMENTS}. This basic structure suggests a bias towards 1- or
2-dimensional data, but it could handle data of up to 6 dimensions;
the \texttt{Z.DATA} array could have as many dimensions as HDS would
support, and the axis structures for the higher dimensions were
labelled -- awkwardly -- from \texttt{T} through to \texttt{W}. An
example layout can be found in \figref{fig:dst}.

Around 1990, the code used by \figaro\ to access DST files was
reworked to handle both DST and NDF files
\citep{1990STARB...6...18S}. Support for NDF did
not use the actual NDF library; instead it used direct HDS calls for
both models, but would use different names for the HDS items it
accessed depending on the data model used in the file. This involved a
significant reworking of the \figaro\ code, but maintained compatibility
with existing DST files. However, it has failed to keep up with
recent changes to NDF, such as support for 64-bit data.

\subsection{Asterix}
\label{app:asterix}

\begin{figure*}[t]
\begin{center}
\includegraphics[width=0.7\textwidth]{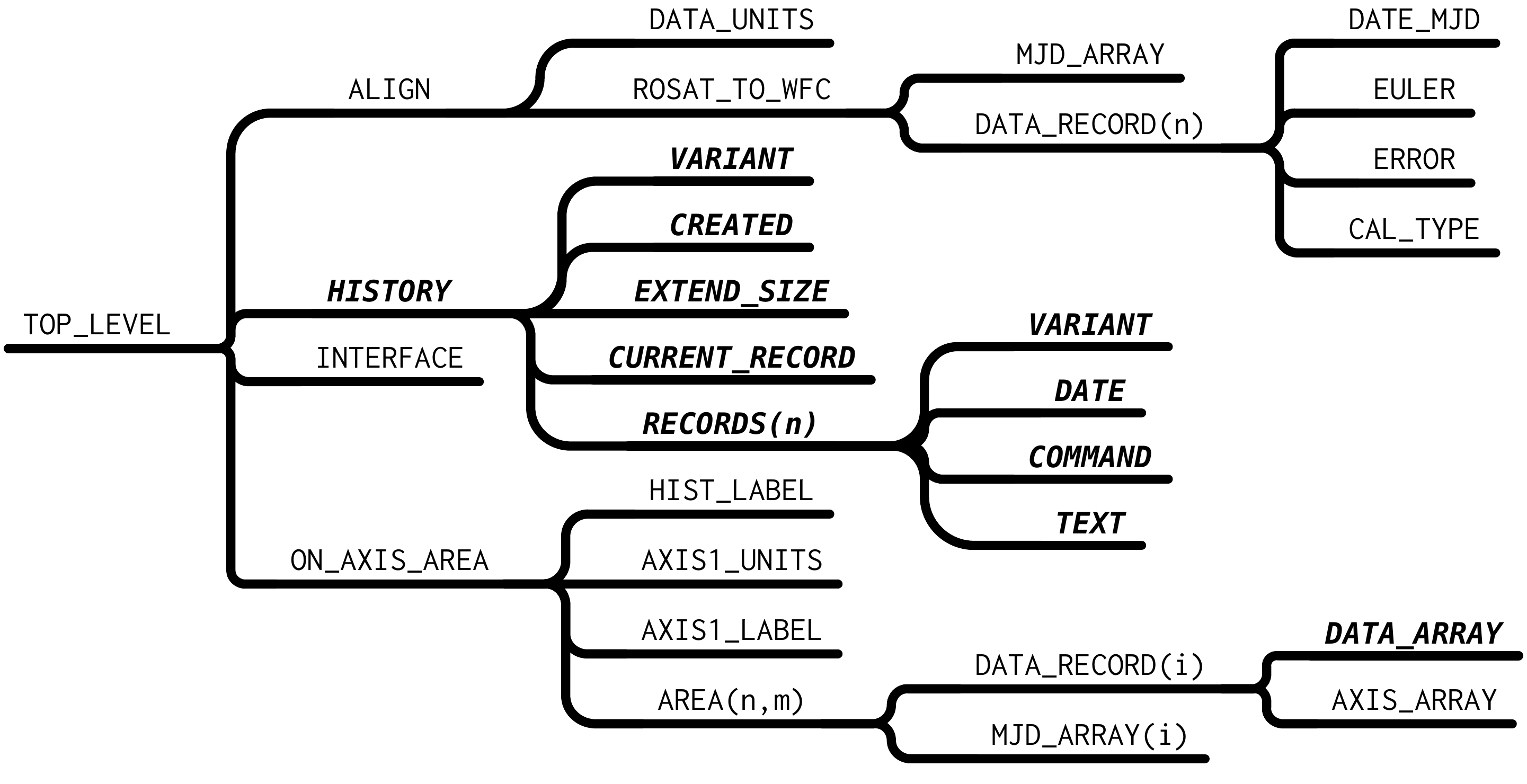}
\end{center}
\caption{Example model of the structure of an \asterix\ HDS
  file. Extensive use is made of hierarchy and the \texttt{HISTORY}
  structure is identical to the NDF standard version. Italicized text
  indicates the components that are shared with NDF.}
\label{fig:asterix}
\end{figure*}

The \asterix\ X-ray data reduction package
\citep[][\ascl{1403.023}]{1987JBIS...40..185P,SUN98,1992STARB...9....3S} used the HDS
format exclusively until the introduction of an abstract data access
interface \citep{1995ASPC...77..199A} which allowed for the use of HDS
and FITS format files. \asterix\ defined many data models designed for
the specific uses of X-ray astronomy, with a particular focus on
photon event lists. An example layout of an alignment file is shown in
\figref{fig:asterix}. The \asterix\ data models were not competing directly with \emph{IMAGE} or
DST but this experience fed directly into the design of NDF.
For example, the \texttt{HISTORY} structure was adopted without change. Once
NDF was available some data models were modified to use features from
NDF such as adopting the \texttt{DATA\_ARRAY} label (see
\figref{fig:duck} for an example).

\subsection{HDS TABLE structure used by FITS2NDF}
\label{app:hdstable}

The \fitstondf\ conversion tool \citep{SUN55} creates an NDF extension
of type \texttt{TABLE} for each FITS TABLE or BINTABLE extension. The
\texttt{TABLE} structure is an extended version of the Giaretta
design.  It comprises a scalar, \texttt{NROWS}, to set the number of
rows in the table, and an array structure \texttt{COLUMNS} also of
type \texttt{COLUMNS}.  The \texttt{COLUMNS} structure contains a
series of \texttt{COLUMN}-type structures, one for each field in the
table. The name of each column comes from the corresponding
\texttt{TTYPE$n$} keyword value.  The conversion does rely on the
column name not being longer than fifteen characters.  A
\texttt{COLUMN} structure has one mandatory component, \texttt{DATA},
an array of values for the field, and can be 2-dimensional if the
field is an array.  Other FITS keywords (\texttt{TFORM$n$},
\texttt{TSCAL$n$}, and \texttt{TFORM$n$}) prescribe the primitive type
of \texttt{DATA}, including expansion of scaled form.  Components to
store the original format, the \texttt{TTYPE$n$} comment, and units
may also be present.

\begin{figure*}[t]
\begin{minipage}{0.5\textwidth}
\begin{quote}
\small
\begin{verbatim}
FITS_EXT_1   <TABLE>       {structure}
  NROWS        <_INTEGER>    4
  COLUMNS      <COLUMNS>     {structure}
    SPORDER      <COLUMN>      {structure}
      COMMENT      <_CHAR*19>    'Spectrum order'
      DATA(4)      <_WORD>       1,2,3,5
      FORMAT       <_CHAR*3>     'I11'

    NELEM        <COLUMN>      {structure}
      COMMENT      <_CHAR*19>    'Number of elements'
      DATA(4)      <_WORD>       1024,1024,1024,1024
      FORMAT       <_CHAR*3>     'I11'

    WAVELENGTH   <COLUMN>      {structure}
      COMMENT      <_CHAR*19>    'Wavelengths of elements'
      DATA(1024,4) <_DOUBLE>     2897.6015206853,
                                 ... 5707.2796545742
      UNITS        <_CHAR*9>     'Angstroms'
      FORMAT       <_CHAR*6>     'G25.16'
\end{verbatim}
\end{quote}
\end{minipage}
\begin{minipage}[t]{0.48\textwidth}
\begin{center}
\includegraphics[width=0.8\textwidth]{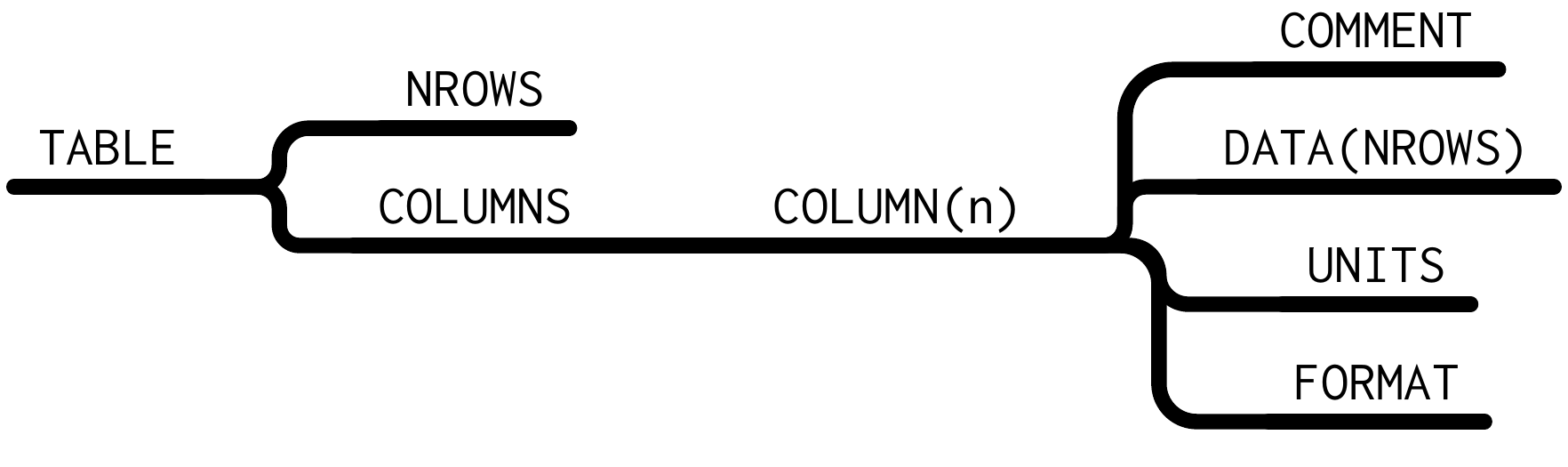}
\end{center}
\end{minipage}
\caption{On the left is a partial dump of the structure of a \texttt{TABLE} structure
within an NDF extension. This structure trace is done with the generic
\texttt{hdstrace} command that lists the content of an arbitrary HDS
structure (similar to the HDF5 \texttt{h5ls} command).
 The name of the NDF extension, in this case \texttt{FITS\_EXT\_1}, can be specified
or take the generic form as above, where the digit counts the
extensions. This example specifies that the table has four rows and three
columns (\texttt{SPORDER}, \texttt{NELEM} and \texttt{WAVELENGTH}). Note
how the structure type (indicated by the angle brackets) indicates how
an application should interpret the contents; in this case the
important types are \texttt{TABLE}, \texttt{COLUMNS} and
\texttt{COLUMN}. On the right is a schematic representing the
structure hierarchy.}
\label{fig:hdstable}
\end{figure*}

An example of part of such a structure is shown in \figref{fig:hdstable}.

\section{Provenance data model}
\label{app:prov}

This appendix describes the data model used to record the ``family tree'' of
ancestor NDFs that were used to create an NDF. Each node in the tree
describes a single NDF, with the root node being the NDF for which
provenance is being recorded. Thus the parents of the root node each
describe one of the NDFs that were used to create the NDF described by the
root node. A node stores the following items of information about the
associated NDF:

\begin{itemize}
\item The path to the NDF within the local file system. This is a blank
string for the root node, since the main NDF may be moved to a new location.
\item The UTC date and time at which the NDF was created.
\item A boolean flag indicating if the NDF is ``hidden'' -- meaning that
the NDF will not be included as an ancestor when provenance is copied
from one NDF to another.
\item A string identifying the command that created the NDF.
\item History information. For the root node, this is just a 4-byte hash code
that represents the contents of the main NDF's \texttt{HISTORY} component. This hash
code is subsequently used to identify the same history information within other
NDFs. For non-root nodes, the history information contains any history
records read from the corresponding ancestor NDF that were not also present
in any of the ancestors direct parents\footnote{This is done to avoid unnecessary
duplication of History records in different nodes.}.
\item Any extra arbitrary information associated with the NDF. There are no conventions on what this extra information
represents.
\item A list of pointers to nodes representing the direct parents of the
NDF.
\end{itemize}

The full tree of nodes is stored on disk in an extension named
\texttt{PROVENANCE} within the main NDF, and is encoded into an array of
integers in order to avoid the overhead of reading and writing complicated
HDS structures.

Each application will normally first create a basic tree for each output
NDF, holding a single root node describing the output NDF. It will then
read the provenance tree from each input NDF and append each one to the
provenance tree of the output NDF, making it a direct parent of the root
node. When the application closes, the final output tree is stored in
the output NDF.

This whole process can be automated by registering handlers with the NDF
library that are called whenever an NDF is opened. The only change that
then needs to be made to an application to enable basic provenance tracking is for the
application to add two calls to mark the start and end of a ``provenance
recording context''. Alternatively, to achieve finer control of which
input NDFs are recorded as parents of each output NDF, it is possible for
an application to handle the reading and writing of provenance trees
itself.

It is possible that a single input NDF may be used many times in the
creation of an output NDF. For instance, if NDF \emph{A} is added to NDF
\emph{B} to create NDF \emph{C}, and \emph{A} is then added to \emph{C} to create
\emph{D}, then \emph{A} (and all its ancestors) would appear twice in the
provenance tree of \emph{D}. To avoid this, whenever a new parent is added to
the root node, each node within the tree of the new parent is compared with
each node already in the tree. If they match, the tree of the new parent
is ``snipped'' at that point to exclude the duplicated node (and all its
parent nodes). This comparison needs to be done carefully since it is
possible for two nodes to include the same path, creator and date, and
yet still refer to different NDFs\footnote{For instance, if an NDF is
created, used once, and then immediately replaced with a new NDF}. For
this reason, the comparison two nodes are considered equal if:

\begin{enumerate}
\item they have the same path, date and creator, and
\item they have the same number of parent nodes, and
\item each pair of corresponding parent nodes are equal
\end{enumerate}

The third requirement above means that two nodes will never be considered
equal if any of the ancestors of the two nodes differ.

\section{NDF structure serialization into FITS}
\label{app:ndf2fits}

The generalized extensions \citep{1988A&AS...73..359G} addition to the
FITS standard made provision for hierarchical data through three
keywords: \texttt{EXTNAME}, \texttt{EXTVER}, and \texttt{EXTLEVEL}.
Indeed \texttt{EXTLEVEL} was added by
Don Wells specifically with Starlink HDS in mind.  However, there were
two concepts missing from these to preserve an NDF structure.  First
and more important is the type of a structure.  Type defines the
semantics and processing rules of an NDF structure.  In
object-oriented parlance it is the class.  Therefore we introduced an
additional keyword \texttt{EXTTYPE} to preserve this information.  Note that
the definition of \texttt{EXTNAME} was somewhat terse and vague in
\citet{1988A&AS...73..359G} being just the \emph{name}, and some FITS
writers have in effect used it as the type (cf.
\secref{sec:name_v_type}).  The second missing
feature was a means to record the shape of a structure.

To map from the hierarchical NDF structure to the flat FITS
serialization array components such as \texttt{VARIANCE} and \texttt{QUALITY} are
written to FITS IMAGE extensions, whose headers retain information
stored in other top-level components such as \texttt{WCS} and \texttt{LABEL}.  Specially
formatted headers may be written to record the \texttt{HISTORY}, which if not
edited, can be read back into NDF \texttt{HISTORY} records.  This is not as
robust as we would like, and a recent formatting change to
\CFITSIO\ \citep[][\ascl{1010.001}]{1999ASPC..172..487P}
temporarily prevented recovery of some records.  Likewise provenance
information, if present, is stored via five keywords, \texttt{PRV[CDIMP]$n$}
for the $n$th NDF. This limits the maximum number of NDFs in the
provenance tree to 9999.  Existing FITS-like headers within the NDF
(\secref{sec:FITSheaders}) are merged, but with NDF information
such as the array shape or WCS superseding the original keyword
values.

NDF extension structures become binary tables when in FITS form.  Each
primitive component within a structure becomes a column in the table
with the appropriate data type and dimension.  Each element of an
array of structures becomes a separate binary table.  A more-compact
storage would be to create a single table for the array structure,
forming a superset of columns, then writing a row for each structure
element, using null values where necessary.  However, the adopted
structure significantly simplified the recursive code.  Also in practice
array structures are small and usually 1-dimensional.

The following header items are set to enable recovery of the hierarchical
structure.

\begin{description}

\item [\texttt{EXTNAME}]
Within a FITS IMAGE extension this is the name of an array component, since
these are in well-defined locations within the NDF. Within a binary
table, which could record arbitrary NDF structures, \texttt{EXTNAME} stores the
dot-separated path within the hierarchical structure.
The path may also include indices to the elements of an array of
structures, written as comma-separated list between parentheses.

A common issue especially for NDF extensions is that the path name is
too long for the 68 characters in a FITS header \texttt{EXTNAME} is set to a
special string \texttt{@EXTNAMEF} that can never be in the component
path.  The full path is written as a long string to keyword EXTNAMEF
using the HEASARC long-string \texttt{CONTINUE}
convention \citep{2007Continue}.

\item [\texttt{EXTTYPE}]
The non-primitive data type of the NDF extension or structure.

\item [\texttt{EXTSHAPE}]
This records the shape of the NDF extension as a comma-separated list.

\item [\texttt{HDUCLAS1}]
Set to \texttt{NDF}.

\item [\texttt{HDUCLAS2}]
The name of the NDF array component.

\end{description}

For the reverse operation, the FITS reader decides whether or not it
knows the semantics of the FITS file to map FITS extensions to NDF
components. Various products from known sources are recognised from
keyword values. In the the case of a former NDF, the reader examines
the \texttt{HDUCLAS1} keyword.  If its value is \texttt{NDF}, the
reader endeavors to recreate the NDF components and extensions from
the keywords listed above, working through the FITS extensions in
order.

For an arbitrary FITS file there is no reason to expect that its data
model maps well to the NDF model.  The default behavior is for the
primary HDU to map to the NDF \texttt{DATA\_ARRAY}, in which blank and
\texttt{NaN} values become the NDF bad value; the WCS headers are used to form
an NDF \texttt{WCS} or \texttt{AXIS} component.  For many cases having
the primary array and metadata is adequate for the conversion.  It's
analogous to the early \emph{IMAGE}-model files operating as NDFs.
However, additional FITS extensions are preserved by default too,
being converted to a series of NDF extensions.  The HDS type of
each NDF extension depends on the \texttt{XTENSION} keyword.  It is
\texttt{NDF} for an IMAGE, and \texttt{TABLE} (\ref{app:hdstable}) for
BINTABLE and TABLE.  The latter representation is often not optimal,
as such extensions can be used to represent many different data
models.  An example is when a binary table holds a series of data
arrays observed at different locations such as from multi-object
spectroscopy.  It is possible with generic HDS tools to manipulate
such data into NDF form.  Where the user knows the mappings between
multi-extension FITS and NDF array components, the FITS reader has a
mechanism for specifying these mappings.

Full details of the conversions between NDF and FITS are available in
\citet{SUN55}.

\end{document}